# Isotope effect, Thermodynamic and Elastic properties of ZrCo and ZrCoH$_3$: An *ab-initio* study

D. Chattaraj[a,*], S.C. Parida[a], Smruti Dash[a], C. Majumder[b]

[a]Product Development Division, [b]Chemistry Division

Bhabha Atomic Research Centre, Trombay, Mumbai 400 085, India

[*]*Corresponding Author*:  D. Chattaraj

*Postal Address*:

Product Development Division,

Bhabha Atomic Research Centre,

Trombay, Mumbai – 400 085,

India

Tel. No. - +91 22 2559 6042

Fax: +91 22 2550 5151

*E-mail*: debchem@barc.gov.in

**Abstract**

The intermetallic compound ZrCo and its hydrides are important materials for their use in hydrogen isotope storage. The dynamical, thermodynamic and elastic properties of ZrCo and its hydrides ZrCoX$_3$ (X= H, D and T) are reported. While the electronic structure calculations are performed using plane wave pseudopotential approach, the effect of isotopes on the vibrational and thermodynamic properties has been demonstrated through frozen phonon approach. The results reveal significant difference between the ZrCoH$_3$ and its isotopic analogs in terms of phonon frequencies and zero point energies. For example, the energy gap between optical and acoustic modes reduces in the order of ZrCoT$_3$ > ZrCoD$_3$ > ZrCoH$_3$. The calculated formation energies of ZrCoX$_3$, including the ZPE, are -146.7, -158.3 and -164.1 kJ/(mole of ZrCoX$_3$) for X = H, D and T, respectively. In addition, the changes in elastic properties of ZrCo upon hydrogenation have also been investigated. The results show that both ZrCo and ZrCoH$_3$ are mechanically stable at ambient pressure. The Debye temperatures of both ZrCo and ZrCoH$_3$ are determined using the calculated elastic moduli.

## 1. Introduction

The ZrCo-$X_2$ (X= H, D and T) systems have gain considerable attention because of its use in the hydrogen isotopes storage in the International Thermonuclear Experimental Reactor (ITER) project [1]. Tritium is one the most important isotopes of hydrogen which is going to be used as fuel in fusion reactor. But, this radioactive isotope is a beta emitter and required to be stored safely in a suitable matrix. The solid state storage of hydrogen isotopes is quite reliable, safe and advantageous compared to gaseous or liquid form of storage [2-4]. Metal hydrides are unique choice as a solid state storage material for tritium [5]. Development and delivery of such systems for tritium are of urgent requirement in the ITER project. Conventionally, uranium is used as getter bed for tritium as it has high absorption capacity at room temperature, low equilibrium pressure (< 1 bar at 550-680 K) which prevents the accidental release of tritium into atmosphere, fast kinetics of hydrogen absorption-desorption and large cyclic life [6]. However, uranium hydride is pyrophoric in nature and uranium is a nuclear material. So investigation for finding out an alternate material for tritium storage is in progress. Several experimental and theoretical studies are reported on the uranium based intermetallics for this purpose [7-9]. Presently, the intermetallic ZrCo have been found to be suitable for the safe storage, supply and recovery of hydrogen isotopes in the ITER [1, 10-14]. ZrCo intermetallic has good hydriding/dehydriding property which can serve as a substitution of uranium [15]. Also it is not pyrophoric and easy to handle as it is not a nuclear material [10, 13]. However, the major drawback of ZrCo is that its absorption-desorption cycle become poor on prolonged thermal cycling [16-18] which is due to the disproportionation of its hydride (ZrCoH$_x$ (x ≤ 3) into the stable hydride phase $ZrH_2$ and the hydrogen non-absorbing phase $ZrCo_2$. Most recently, several experimental studies are reported on the doping of a third element into ZrCo intermetallic to improve its cyclic life stability [19, 20].

Experimental investigation of thermo-physical, vibrational and mechanical properties of metal and alloy tritides are difficult compare to hydrides and deuterides because tritium is radioactive in nature. A special experimental facility is required for the safe handling of tritides. The computational techniques based on first principles method are helpful for this purpose. The theoretically calculated thermo-physical properties of radioactive tritides will be helpful for predicting the behavior of the material where there is lack of experimental facility for handling radioactivity. The computed properties will also serve as supportive data for further experimental findings. In this context, the structural, dynamical, thermodynamic and elastic properties of ZrCo and its hydrides $ZrCoX_3$ (X= H, D and T) are calculated using the DFT based *ab-initio* method.

Very few literatures on theoretical studies of ZrCo-hydrogen system [21-23] are available though there are plenty of experimental data on this system. Gupta [22] has investigated that the Zr-H bonding contribution plays a crucial role in the stability of the hydride $ZrCoH_3$ and also has important role in the hydrogen site occupancy. In our previous study, the structural, electronic and thermodynamic properties of the ZrCo and its hydride $ZrCoH_3$ are reported [23]. The ground state properties like equilibrium lattice constants, bulk modulus and enthalpy of formation of ZrCo and $ZrCoH_3$ have been determined by optimizing the atomic and electronic structure of the compounds. The nature of chemical bonding in ZrCo and $ZrCoH_3$ has been depicted in terms of electronic density of states spectrum and charge density contour. The scope of studying isotope effect, vibrational, thermodynamic and elastic properties of ZrCo and $ZrCoH_3$ is fulfilled here. Recently, the isotope effect of $ZrX_2$ compounds is reported using first principles method [24]. In that study, the isotope effect of $ZrX_2$ (X= H, D and T) compounds are depicted in terms variation of phonon frequencies and zero point energies. Li *etal*. [25] have reported the structural, vibrational and thermodynamic properties of ZrCo by first principles method and density-functional perturbation theory (DFPT). The phonon frequency (ω) at the Brillouin zone center, Phonon dispersion curve and phonon density of state for ZrCo have been determined. Zero point energy and the phonon contribution to the thermodynamic properties such as Helmholtz free energy, internal energy, entropy and constant-volume specific heat of ZrCo are calculated from 300 to 1000 K within the harmonic approximation [26]. Regarding the elastic properties of ZrCo and its hydride, Agosta etal. [27] have experimentally investigated the variation of elastic properties of ZrCo as a function of temperature using ultrasonic pulse echo method. Very few literatures are available on the elastic properties of ZrCo and $ZrCoH_3$.

As the electronic properties of all $ZrCoX_3$ (X= H, D and T) are similar, it is of interest to investigate their vibrational and thermodynamic properties. In particular this study focuses on the isotope effect of the $ZrCoX_3$ (X= H, D and T) compounds through phonon frequencies and thermodynamic parameters. These results provide useful, reliable and important informations about the $ZrCoX_3$ systems which will be complementary to the experimental values. The calculated elastic properties of ZrCo and its hydride ZrCoH$_3$ may help as important informations for the design of the tritium storage bed.

## 2. Computational details

All the present calculations are performed using the plane wave-pseudopotential method under the framework of density functional theory as implemented in the Vienna *ab-initio* simulation package (VASP) [28-30]. The electron-ion interaction and the exchange correlation energy are described under the projector-augmented wave (PAW) [31,32] method and the generalized gradient approximation (GGA) of Perdew-Burke-Ernzerhof (PBE) [33], respectively. The valence electronic configuration of Zr, Co and H are set to $5s^14d^3$, $4s^13d^8$ and $1s^1$, respectively. The energy cut off for the plane wave basis set is fixed at 500 eV. The ionic optimization is carried out using the conjugate gradient scheme and the forces on each ion was minimized upto 5meV/Å [34,35]. The k-point sampling in the Brillouin Zone (BZ) has been treated with the Monhorst-Pack scheme [36], using a 4x4x4 k-mesh. Total energies of each relaxed structure using the linear tetrahedron method with Blöchl corrections are subsequently calculated in order to eliminate any broadening-related uncertainty in the energies [37]. To begin with the dynamical calculations, the lattice parameters of ZrCo and ZrCoH$_3$ have been optimized using VASP code and the optimized structures are used for phonon calculation.

The phonon frequencies of ZrCo and ZrCoX$_3$ (X=H, D and T) are calculated by the PHONON program [38] using the forces based on the VASP package. A 3x3x3 supercell of ZrCo containing total 54 atoms and a 3x1x2 supercell of ZrCoX$_3$ (X=H, D and T) containing 120 atoms have been used for the phonon calculations. A small displacement of 0.02 Å have been given to the atoms present in the supercell of ZrCo and ZrCoX$_3$ (X=H, D and T) compounds. The phonon dispersion curves and temperature dependent thermodynamic functions of these compounds are obtained by using the calculated phonon frequencies. The temperature-dependent thermodynamic functions of a crystal, such as the internal energy ($E$), entropy ($S$), Helmholtz free energy ($F$) and constant volume heat capacity ($C_V$) can be calculated from their phonon density of states as a function of phonon frequencies. In the present study, the phonon contribution to Helmholtz free energy $F$, internal energy $E$, entropy $S$ and constant volume specific heat $C_V$, at temperature $T$ are calculated within the harmonic approximation using the following formulas [39]:

$$F = 3nNk_BT \int_0^{\omega_{max}} \ln\left\{2\sinh\frac{\hbar\omega}{2k_BT}\right\} g(\omega)d\omega \tag{1}$$

$$E = 3nN\frac{\hbar}{2} \int_0^{\omega_{max}} \omega \coth\left(\frac{\hbar\omega}{2k_BT}\right) g(\omega)d\omega \tag{2}$$

$$S = 3nNk_B \int_0^{\omega_{max}} \left[\frac{\hbar\omega}{2k_BT} \coth\frac{\hbar\omega}{2k_BT} - \ln\left\{2\sinh\frac{\hbar\omega}{2k_BT}\right\}\right] g(\omega)d\omega \tag{3}$$

$$C_V = 3nNk_B \int_0^{\omega_{max}} \left(\frac{\hbar\omega}{2k_BT}\right)^2 \csc h^2\left(\frac{\hbar\omega}{2k_BT}\right) g(\omega) d\omega \tag{4}$$

$F$ and the $E$ at zero temperature represents the zero point energy, which can be calculated from the expression as $F_0 = E_0 = 3nN \int_0^{\omega_{max}} \left(\frac{\hbar\omega}{2}\right) g(\omega)d\omega$, where $n$ is the number of atoms per unit cell, $N$ is the number of unit cells, $\omega$ is the phonon frequencies, $\omega_{max}$ is the maximum phonon frequency, and g($\omega$) is the normalized phonon density of states with $\int_0^{\omega_{max}} g(\omega)d\omega = 1$. The total energy of hydrogen molecule (H$_2$) and zero point energy of X$_2$ (X= H$_2$, D$_2$ and T$_2$) molecules are calculated using DFT which is described earlier [24]. The elastic properties of ZrCo and its hydride ZrCoH$_3$ were also calculated using an efficient stress-strain method [40] implemented in VASP.

## 3. Results and discussion

### 3.1 Structural properties

The crystal structure of ZrCo is CsCl-type cubic (bcc) with lattice parameter $a$ = 3.196 Å [41, 42] as shown in fig. 1(a). In ZrCo, the Zr atom occupies 1$a$ (0, 0, 0) and Co atom occupies 1$b$ (0.5, 0.5, 0.5) Wyckoff site. The hydride ZrCoH$_3$ favors a simple orthorhombic ZrNiH$_3$-type crystal structure as shown in fig. 1(b) with the room temperature lattice parameters listed in table 1 [43, 44]. The crystal structure shown in fig. 1(b) contains two unit cells of ZrCoH$_3$. To obtain the ground state structural parameters, the ionic and electronic structure of the ZrCo and ZrCoH$_3$ have been optimized by varying the lattice parameters. The ground state crystal structures data and the optimized lattice parameters of ZrCo and ZrCoH$_3$ are summarized in table 1. The lattice parameters are found to be within ± 1% accuracy from the experimental data. The good agreement between calculated lattice parameters and the experimentally reported values establishes the accuracy and reliability of the present computational method. The calculated ground state structure of ZrCo and ZrCoH$_3$ are considered for the calculation of vibrational and elastic properties.

*3.2 Vibrational properties*

According to lattice vibration theory [45], vibrational frequency $\omega$ can be expressed as a function of both direction and magnitude of wave vector $q$ using the dispersion relation:

$$\omega = \omega_j(q) \tag{5}$$

The subscript j is the branch index. Generally, a crystal lattice with n atoms per unit cell has 3n branches, three of which are acoustic modes and the remainders are optical modes. The lattice vibration mode with $q \approx 0$ plays an important role for Raman scattering and infrared absorption [46]. So, the vibrational frequency with $q = 0$, *i.e.* at the centre $\Gamma$ point of the first Brillouin zone, is called as normal mode of vibration. The crystal structure of ZrCo contains 2 atoms per unit cell, so there are six normal modes of vibrations, which includes three low frequency acoustic modes and three high frequency optical modes. As the ZrCoX$_3$ (X= H, D and T) compounds contains two unit cell having total number of 10 atoms, there are 30 normal vibrational modes among which three are acoustic and the remainder are optical modes. The light atom H has larger displacement amplitude which corresponds to high frequency optical modes and heavy atoms Zr and Co corresponds to low frequency optical modes. According to group theory [47], the irreducible representations of normal modes at Brillouin zone centre ($\Gamma$ point) for ZrCo can be expressed as

$\Gamma_{aco} = 3T_{1u}(IR), \quad \Omega_{acu} = 0.012$ THz

$\Gamma_{opt} = 3T_{1u}(IR), \quad \Omega_{opt} = 5.318$ THz

Similarly, the phonon frequencies at Brillouin zone centre ($\Gamma$ point), the IR and Raman active modes of ZrCoX$_3$ (X= H, D and T) are given in table 2.

The (IR) and (R) stand for infra red active and Raman active modes respectively; subscript $u$ and $g$ represents antisymmetric and symmetric modes respectively with respect to the center of inversion.

The phonon dispersion curves show how the phonon energy depends on the q-vectors along the high symmetry directions in the Brillouin zone. This can be compared with the experimental graph obtained from the neutron scattering experiments on single crystals. The phonon dispersion curves at 0 K for ZrCo and $ZrCoX_3$ (X= H, D, T) have been obtained by plotting vibrational frequencies along the high symmetry directions, as shown in Figs. 2-5. For ZrCo intermetallic, all the acoustic and optical phonon frequencies are positive, while for $ZrCoX_3$ (X= H, D, T) few acoustic frequencies are negative. This indicates that ZrCo is dynamically stable whereas $ZrCoX_3$ are not. The phonon dispersion curves of $ZrCoH_3$ and its analogue are almost similar but there are some distinct features in which they actually differ. The mass difference between the constituent elements of the compound significantly affects the maximum and minimum values of the acoustic and optical branches, and a clear gap is formed between them, as can be seen in Figs. 2-5. The gap between the high frequency optical modes and low frequency acoustic modes decreases by isotopic substitution in $ZrCoX_3$ from ~22 THz ($ZrCoH_3$) to ~16 THz ($ZrCoD_3$) and ~11 THz ($ZrCoT_3$). The higher frequency peaks come closer to Zr and Co peaks as we substitute H of $ZrCoH_3$ with the isotopes D and T which is attributed to the increase in mass.

*3.3 Thermodynamic properties*

The calculated zero point energies (ZPE) are 5.01, 43.7, 32.1 and 26.2 kJ/mol for the ZrCo, $ZrCoH_3$, $ZrCoD_3$ and $ZrCoT_3$, respectively. The enthalpy of formation of ZrCo (at 0 K) changes from -55.9 kJ/mol to -51.0 kJ/mol after ZPE correction. The enthalpy of formation ($\Delta_f H$) at 0 K for $ZrCoH_3$ is -190.3 kJ/mol without considering ZPE. After including the ZPE correction, the heat of formation of $ZrCoH_3$ changes from -190.3 to -146.7 kJ/mol. Similarly, ZPE corrected $\Delta_f H$ at 0 K for $ZrCoD_3$ and $ZrCoT_3$ are -158.3 and -164.1 kJ/mol, respectively. It is interesting to note that, although $ZrCoH_3$ and its isotopic analogues $ZrCoD_3$ and $ZrCoT_3$ have the same crystal and electronic structure, $ZrCoT_3$ and $ZrCoD_3$ are more stable than $ZrCoH_3$. The variation of *F*, *E*, *S* and $C_v$ are shown in Fig. 6(a-d) upto 600 K, which is below the decomposition temperature of $ZrCoX_3$ (X= H, D, and T). Fig. 6(a) shows that the Helmholtz free energy (*F*) for all three hydrides decreases gradually with increase in temperature. In contrast, the internal energy *E* and entropy *S* increases with increase in temperature as shown in Fig. 6(b) and 6(c).

Heat capacity of a solid is an important thermodynamic parameter which depicts the behavior of that material in different thermal conditions. We have also calculated the heat capacities of ZrCo and ZrCoX$_3$ (X = H, D, and T). The temperature dependent heat capacity ($C_v$) of ZrCo and ZrCoX$_3$ (X= H, D, T) is shown in Fig. 6(d). It is observed that the $C_v$ of ZrCo increases rapidly upto 100 K and gradually attains a constant value of ~ 49 J/mol K. The calculated specific heat of ZrCo is found to be 2.8 x 10$^{-3}$ J/mol K. This is in good agreement with the experimental value of 2.45 x 10$^{-3}$ J/ mol K at 3 K [27]. It is also seen that at low temperature, upto 300 K, the heat capacities of ZrCoX$_3$ (X = H, D, and T) increase rapidly with increase in temperature and thereafter increases slowly up to 600 K, and attain the saturation value which is known as Dulong-Petit classical limit. From the fig. 6(d) it is seen that upto 100 K, the variation of $C_v$ versus $T$ curve follows the same trend for all ZrCoX$_3$ compounds, but above 100 K it shows different trend: $C_v$ (ZrCoT$_3$) > $C_v$ (ZrCoD$_3$) > $C_v$ (ZrCoH$_3$). The heat capacities of hydrides are higher compare to its precursor alloy ZrCo. It is due to the contribution of high frequency optical modes of X$_2$ (X= H, D and T) to the $C_v$ of hydrides. In addition, the heat capacity plot shows a broad hump around 100 K for all three compounds. At low temperature, below 100 K, the acoustic modes, in which the Zr, Co and H atoms are vibrating 'in-phase', dominate the heat capacity contribution. At higher temperature, above 100 K, the optical modes of vibration, in which the light H atoms vibrate against the heavy and almost stationary Zr and Co atoms, dominates the heat capacity. Hence, around 100 K, a transition from predominant acoustic vibration to optical vibration cause a change in heat capacity function leading to a broad hump.

## 4. Elastic Properties

The mechanical properties of solids are very important for various reasons, particularly, it depicts the behavior of a material under different stress and strain conditions [48]. The structural and mechanical stability, deformations, phase transformations, bonding characteristics, melting point etc. can be described by elastic stiffness parameters. The elastic constants of a solid are also related to mechanical properties such as bulk modulus, shear modulus, Young's modulus, poison's ratio and elastic anisotropy. The elastic constants are also useful to determine Debye temperature. Investigation of elastic contribution to the hydrogen-hydrogen interaction energy in metal hydrides [49] requires adequate knowledge of elastic constants. To investigate the mechanical stability of ZrCo and its hydride ZrCoH$_3$, a set of zero pressure elastic constants have been determined from the stress-strain approach [50] as implemented in the VASP code. The calculated single crystal elastic constants of ZrCo and its hydride ZrCoH$_3$ are listed in table 3.

For body centered cubic (bcc) symmetry, there are three independent elastic constants, namely, $C_{11}$, $C_{12}$ and $C_{44}$. The conditions for mechanical stability of cubic crystals are: $C_{11}-C_{12} > 0$, $C_{11} > 0$, $C_{44} > 0$, $C_{11}+2C_{12} > 0$, $C_{12} < B < C_{11}$ [51]. The calculated elastic constants ($C_{ij}$) of ZrCo alloy satisfy all of these conditions, hence it can be said that ZrCo is mechanically stable. Similarly, for mechanically stable orthorhombic crystal, nine independent elastic constants $C_{ij}$ should satisfy the following criteria [45]:

$$C_{11} > 0, C_{22} > 0, C_{33} > 0, C_{44} > 0, C_{55} > 0, C_{66} > 0 \tag{6}$$

$$[C_{11} + C_{22} + C_{33} + 2(C_{12} + C_{13} + C_{23})] > 0 \tag{7}$$

$$(C_{11} + C_{22} - 2C_{12}) > 0 \tag{8}$$

$$(C_{11} + C_{33} + 2C_{13}) > 0 \tag{9}$$

$$(C_{22} + C_{33} + 2C_{23}) > 0 \tag{10}$$

In this study, the calculated elastic constants $C_{ij}$ of orthorhombic $ZrCoH_3$ satisfy above conditions. Hence, $ZrCoH_3$ is mechanically stable at ambient pressure. In ZrCo, $C_{11} > C_{12} > C_{44}$, indicates that the bonding strength is strongest in the (100) direction. In case of $ZrCoH_3$, it is seen that $C_{11}$ and $C_{33}$ have, almost, same value, which indicates that the atomic bonding between nearest neighbors along the (100) and (001) planes, have the same strength. However, $C_{22} > C_{33}$ for $ZrCoH_3$ implies that the atomic bonds along the (010) planes between nearest neighbors are stronger than those along the (001) plane. This may be attributed to the bond formation between Co and H atoms in the (010) plane. In both ZrCo and its hydride $ZrCoH_3$, $C_{11} > C_{44}$ suggests that [100](100) shear is easier than [100](010) shear. Comparison of the elastic constants for the precursor alloy and its hydride indicates that the hydride is more resistant to both compressions in the a- and c-directions and especially to shear deformation, as $C_{11}$, $C_{12}$ and $C_{44}$ of $ZrCoH_3$ are higher than that of ZrCo. For ZrCo and $ZrCoH_3$, $C_{11} > C_{44}$, So deformation in the perpendicular direction to a-axis is easier than along c-axis. The shear elastic constants $C_{44}$, $C_{55}$ and $C_{66}$ are indicative of resistance to shear deformation whereas $C_{11}$, $C_{22}$ and $C_{33}$ are related to the unidirectional compression along the principal crystallographic directions. The calculated result shows that $C_{44}$, $C_{55}$ and $C_{66}$ values are more than 50% lower than that of $C_{11}$, $C_{22}$ and $C_{33}$. This depicts that $ZrCoH_3$ has weak resistance to shear deformation compared to that of the unidirectional compression.

The elastic moduli of polycrystalline materials are generally calculated by two approximations, namely, Voigt [52] and Reuss [53] in which uniform strain or stress are assumed throughout the polycrystal. Later, Hill [54] proposed that the actual elastic moduli is the arithmetic mean of the Voigt and Reuss values, which is known as the Voigt-Reuss-Hill (VRH) value. The details of the calculations for the bulk and shear moduli is given in somewhere else [55, 56] and therefore not recalled here. The Young's modulus and the Poisson's ratio can be obtained from the bulk and shear moduli [56]. The Cauchy pressure σ, Zener anaisotropy factor A, Poisson's ratio ν and Young's modulus Y, which are important elastic parameters, are calculated using the following relations [57]:

$$\sigma = (C_{12} - C_{44}) \qquad (11)$$

$$A = \frac{2C_{44}}{C_{11} - C_{12}} \qquad (12)$$

$$\nu = \frac{3B - 2G}{2(3B + G)} \qquad (13)$$

$$Y = \frac{9GB}{G + 3B} \qquad (14)$$

where $G = (G_V + G_R)/2$ is the isotropic shear modulus, $G_V$ is Voigt's shear modulus corresponding to the upper bound of G values, and $G_R$ is Reuss's shear modulus corresponding to the lower bound of G values.

The Zener anaisotropy, bulk modulus, shear modulus, Young's modulus and Poisson's ratio have been estimated from the calculated single crystal elastic constants, and are given in table 4. Pettifor [58] and Johnson [59] suggested that ductile or brittle behavior of a material can be predicted by the Cauchy Pressure ($C_{12}$-$C_{44}$). For ductile material, Cauchy pressure is positive, while for brittle material, it is negative. For ZrCo, Cauchy pressure was found to be positive which indicates that ZrCo is ductile in nature. The Zener anaisotropy factor (A) represents the degree of elastic anaisotropy in solids. The *A* takes the value of 1 for a completely isotropic material. If the value of *A* is smaller or greater than unity it shows the degree of elastic anaisotropy. The calculated Zener anisotropy factor for ZrCo is 1.80 at 0 GPa, which indicates that the compounds are entirely anisotropic. The shear anisotropic factors of different Miller planes express the degree of anaisotropy in atomic bonding in those planes. The shear anisotropic factors are expressed as follows:

$$A_1 = \frac{4C_{44}}{C_{11}+C_{33}-2C_{13}} \quad \text{for the 100 plane} \tag{15}$$

$$A_2 = \frac{4C_{55}}{C_{22}+C_{33}-2C_{23}} \quad \text{for the 010 plane} \tag{16}$$

$$A_3 = \frac{4C_{66}}{C_{11}+C_{22}-2C_{12}} \quad \text{for the 001 plane} \tag{17}$$

The calculated $A_1$, $A_2$ and $A_3$ of $ZrCoH_3$ are 1.64, 1.0 and 1.25 respectively. For an isotropic crystal, the factors $A_1$, $A_2$ and $A_3$ must be equal to 1, while any value smaller or greater than 1 represents varying degree of anaisotropy. The calculated results indicate that $ZrCoH_3$ is isotropic in (010) plane while anisotropic in (100) and (001) planes.

The average bond strengths between the atoms in a crystal can be predicted by knowing its bulk modulus [60]. The calculated bulk modulus of $ZrCoH_3$ is 162 GPa which is higher than that of ZrCo (146 GPa). This suggests that hydrogenation of ZrCo resulting $ZrCoH_3$ increases the fracture strength of ZrCo. The resistance of a material to size and shape change can be measured by the bulk modulus B and shear modulus G, respectively. The calculated results show that $ZrCoH_3$ is more resistant to size and shape changes than its precursor alloy ZrCo. Young's modulus indicates the resistance against uniaxial tensions and is indicative of stiffer material. Accordingly, the high Young modulus value of $ZrCoH_3$ compare to ZrCo suggests that this hydride is stiffer than its precursor alloy. Poisson's ratio is the ratio of transverse contraction strain to longitudinal extension strain under a stretching force. It is also related to the bonding properties of materials. Poisson's ratio varies in different materials depending on the nature of bonding present in those materials. As for covalent materials, the value of ν is small (typically ν = 0.1) whereas for ionic and metallic materials, the typical value of ν are 0.25 and 0.33, respectively [61]. The Poisson's ratio of ZrCo is 0.34, which indicate that it is metallic in nature. The lower limit and upper limit of Poisson's ratio ν are given 0.25 and 0.5, respectively, for central forces in solids [62]. In solids, most of the measured values fall in the range of 0.28 - 0.42. Our calculated values show that the interatomic forces in the ZrCo and $ZrCoH_3$ are predominantly central in nature. Pugh suggested that the ratio of bulk (B) to shear modulus (G) plays an important role to predict the ductile or brittle behavior of a solid material [63]. According to the Pugh condition, if B/G > 1.75, then ductile behavior is predicted, otherwise the material behaves in a brittle manner. The ratio for ZrCo is larger than 1.75, and the results suggest that the ZrCo phase is ductile in nature which is supported by the calculated Cauchy pressure in this study. It has recently been reported that elastic properties, specially ductility, bulk to shear modulus ratio (B/G), Poisson's ratio (ν) etc., are related to the initial hydriding mechanism [64]. Here we have compared the ductile behavior of few hydrogen isotope storage materials in terms of B/G ratio and Poisson's ratio (ν) and shown in table 5. It is seen from table 4 that Pd has higher B/G value (3.97) compare to that of U (1.64) and $U_2Ti$ (1.21) whereas ZrCo takes up the value of 3.24. According to Pugh's condition Pd and ZrCo are more ductile (B/G > 1.75) while U and $U_2Ti$ are brittle in nature (B/G < 1.75). It is also interesting to note that Pd has higher ν value compare to U and $U_2Ti$ while ZrCo has an intermediate value. A disadvantage of uranium to be used as hydrogen isotope storage material is that it pulverizes into very fine powder [65] on repeated hydriding/dehydriding whereas Pd does not. The higher B/G and ν values of Pd compare to U and $U_2Ti$ may be the reason for that. From this discussion, it can be suggested that if B/G and ν value of ZrCo is increased by any means, let's say by introducing a third element in ZrCo intermetallic, its powdering property can be improved and it can be used as a better material for hydrogen isotope storage.

The elastic moduli of a solid are also related to the thermal properties of solid through the Debye theory. The Debye temperature ($\theta_D$) is related to many important physical properties such as specific heat, elastic constants, and melting point [66]. The Debye temperature and mean, transverse, longitudinal sound velocities have been calculated using the following well-known relations [67]:

$$\theta_D = \frac{\hbar}{k}\left[\frac{3n}{4\pi}\left(\frac{N_A\rho}{M}\right)\right]^{1/3} v_m \qquad (18)$$

where $\hbar$ is the Planck's constant, $k$ is Boltzman's constant, $N_A$ is Avogadro's number, $n$ is the number of atoms per formula unit, $\rho$ is the density.

Mean ($v_m$), transverse ($v_t$) and longitudinal ($v_l$) sound velocities are given, respectively, as

$$v_m = \left[\frac{1}{3}\left(\frac{2}{v_t^3} + \frac{1}{v_l^3}\right)\right]^{1/3} \qquad (19)$$

$$v_t = \left(\frac{G}{\rho}\right)^{1/2} \qquad (20)$$

$$v_l = \left(\frac{3B+4G}{3\rho}\right)^{1/2} \qquad (21)$$

The calculated elastic constant can be used to determine the melting temperature $T_m$ of solids. The melting temperature of ($T_m$) has been estimated using an empirical relation [68],

$$T_m = 553 + (591/MBar)C_{11} \pm 300K \tag{22}$$

Using the equation (22), the melting point of ZrCo is calculated to be 1581.3 ± 300 K which is agreement with the experimental value melting point (1622 K) of ZrCo [69].

All calculated quantities from Eq. (18 - 22) are listed in table 6. The calculated value of Debye temperature of ZrCo agrees well with the experimental value of 281.3 K [27]. It is seen from the table 4 that, Debye temperature ($\theta_D$) of ZrCo is lower than $ZrCoH_3$. As Debye temperature of a solid represents the interatomic force, the high $\theta_D$ value of $ZrCoH_3$ compare to its ZrCo indicates that $ZrCoH_3$ has stronger bonds than ZrCo.

## 5. Conclusions

In this study, the first principles calculations of crystal structure, vibrational and thermodynamic properties of ZrCo and $ZrCoX_3$ (X= H, D, T) are performed. The phonon frequencies phonon dispersion curves have been obtained using the frozen phonon approach. The Raman and infrared active modes present in the compounds of our study are also been assigned. Using the calculated phonon frequencies, the thermodynamic functions are determined with the harmonic approximation. The ZPE corrected enthalpies of formation ($\Delta_f H$ at 0 K) are obtained for the compounds $ZrCoX_3$ (X= H, D, T) and those suggest that $ZrCoT_3$ and $ZrCoD_3$ are more stable than $ZrCoH_3$. The isotope effect is also reflected on the thermodynamic properties of the compounds under study. The ZPE corrected enthalpy of formation data at 0 K coupled with the high temperature data of ZrCoX3 (X= H, D, T) will help us to calculate the equilibrium dissociation pressure of X2 (X= H, D, T) for practical applications. Along with the dynamical and thermodynamic properties, the effect on elastic properties upon hydrogenation of ZrCo is also explored in this study. Both ZrCo and its hydride $ZrCoH_3$ are found to be mechanically stable. On the basis of elastic properties it is also suggested that ZrCo is a better material compare to U and $U_2Ti$ for the purpose of tritium storage.

**Acknowledgements**

The authors are thankful to Dr. K.L. Ramakumar, Director, Radiochemistry and Isotope Group, Bhabha Atomic Research Centre (BARC) and Dr. S.K. Mukerjee, Head, Product Development Division, BARC for their interest and encouragement during progress of this work. The authors are also thankful to the members of the Computer Division, BARC, for their kind cooperation during this work.

**Table Captions**

Table 1: Optimized lattice constants along with the experimental values for ZrCo and ZrCoH$_3$.

Table 2: Phonon frequency at the $\Gamma$ point of ZrCoX$_3$ (X= H, D and T).

Table 3 Calculated elastic constants C$_{ij}$ (in GPa) of ZrCo and ZrCoH$_3$.

Table 4 Calculated elastic moduli (in GPa), Zenar anisotropy (A), Poisson's ratio ($\nu$), ratio of B/G of ZrCo and ZrCoH$_3$.

Table 5 Comparison of bulk modulus (B), shear modulus (G), B/G ratio and Poisson's ratio for different hydrogen isotope storage materials.

Table 6 Calculated sound velocities (m/s) and Debye temperature (K) of ZrCo and ZrCoH$_3$.

| System | Crystal structure | Space group | Calculated (0 K) | Experimental (298 K) [Ref.] |
|---|---|---|---|---|
| ZrCo | bcc | *Pm-3m* | $a$ (Å) = 3.181 | 3.196 [41,42] |
| ZrCoH$_3$ | orthorhombic | *Cmcm* | $a$ (Å) 3.531<br>$b$ (Å) 10.395<br>$c$ (Å) 4.311 | 3.527 [43,44]<br>10.463<br>4.343 |

Table 1

| ZrCoX$_3$ (X = H, D and T) | | | | | | | | | | |
|---|---|---|---|---|---|---|---|---|---|---|
| *IR* | | | | *Raman* | | | | *Silent* | | |
| | H | D | T | | H | D | T | | H | D | T |
| Mode | Ω | Ω | Ω | Mode | Ω | Ω | Ω | Mode | Ω | Ω | Ω |
| B2u | -0.598 | -0.592 | 0.196 | B1g | 4.158 | 4.141 | 4.153 | Au | 30.682 | 21.696 | 18.466 |
| B1u | -0.115 | -0.114 | -0.108 | B1g | 4.343 | 4.331 | 4.505 | | | | |
| B3u | 0.198 | 0.196 | 0.209 | Ag | 4.413 | 4.401 | 4.383 | | | | |
| B3u | 5.000 | 4.979 | 5.311 | B3g | 4.578 | 4.557 | 4.650 | | | | |
| B1u | 5.018 | 5.001 | 5.220 | B3g | 6.392 | 6.364 | 6.617 | | | | |
| B2u | 5.055 | 5.042 | 4.838 | Ag | 6.756 | 6.730 | 6.931 | | | | |
| B3u | 28.957 | 20.669 | 17.393 | B1g | 29.262 | 20.767 | 17.295 | | | | |
| B1u | 28.967 | 20.508 | 16.533 | B3g | 30.050 | 21.300 | 17.582 | | | | |
| B1u | 29.147 | 20.734 | 17.505 | B3g | 30.296 | 21.447 | 17.849 | | | | |
| B2u | 30.021 | 21.309 | 18.512 | Ag | 30.615 | 21.686 | 18.609 | | | | |
| B3u | 31.713 | 22.521 | 18.793 | B2g | 32.566 | 23.028 | 19.764 | | | | |
| B2u | 32.940 | 23.397 | 18.640 | Ag | 32.912 | 23.283 | 19.109 | | | | |
| B1u | 41.494 | 29.517 | 19.721 | B1g | 33.759 | 23.950 | 20.254 | | | | |
| B2u | 43.462 | 30.876 | 21.384 | B3g | 42.756 | 30.394 | 20.391 | | | | |
| | | | | Ag | 44.569 | 31.647 | 21.911 | | | | |

Table 2

| System | | C11 | C12 | C13 | C22 | C23 | C33 | C44 | C55 | C66 |
|---|---|---|---|---|---|---|---|---|---|---|
| ZrCo | This Study | 174 | 134 | - | - | - | - | 36 | - | - |
| ZrCoH$_3$ | This Study | 243 | 124 | 159 | 257 | 93 | 241 | 68 | 78 | 79 |

Table 3

| System | | $B$ | $G$ | $L$ | $Y$ | $A$ | $v$ | $B/G$ |
|---|---|---|---|---|---|---|---|---|
| ZrCo | This Study | 145.7 | 45.0 | 182.1 | 122.4 | 1.80 | 0.36 | 3.24 |
| | [a]Reported | 145.3 | 52.8 | - | 141 | - | 0.34 | 2.75 |
| $ZrCoH_3$ | This Study | 162 | 63 | 246 | 168 | $A1 = 1.64$ | 0.33 | 2.57 |
| | | | | | | $A2 = 1.0$ | | |
| | | | | | | $A3 = 1.25$ | | |

[a]ref [27]

Table 4

| Material | | B (GPa) | G (GPa) | B/G | ν |
|---|---|---|---|---|---|
| U | [b]Expt. | 119.1 | 72.6 | 1.64 | 0.25 |
| U$_2$Ti | [b]Expt. | 124.5 | 103.2 | 1.21 | 0.18 |
| ZrCo | [c]Expt. | 140 | 49.2 | 2.85 | 0.34 |
| ZrCo | Theory (This study) | 145.7 | 45.0 | 3.24 | 0.36 |
| Pd | [d]Expt. | 190.0 | 47.9 | 3.97 | 0.384 |

[b]Ref [70]
[c]Ref [27]
[d]Ref [71]

Table 5

| System | | $V_l$ (m/s) | $V_t$ (m/s) | $V_m$ (m/s) | $\theta_D$ (K) | $T_m$ (K) |
|---|---|---|---|---|---|---|
| ZrCo | This Study | 4785 | 2433 | 2785 | 259.3 | 1581.3 ± 300 |
| | [a]Reported | 5338 | 2640 | 2960 | 281.3 | - |
| ZrCoH$_3$ | This Study | 6191 | 3139 | 3593 | 315.8 | - |

[a]Ref [27]

Table 6

**Figure Captions**

Fig. 1(a) and 1(b) Crystal structure of ZrCo and ZrCoH$_3$.

Fig. 2 Calculated Phonon dispersion graph of ZrCo.

Fig. 3 Calculated Phonon dispersion graph of ZrCoH$_3$.

Fig. 4 Calculated Phonon dispersion graph of ZrCoD$_3$.

Fig. 5 Calculated Phonon dispersion graph of ZrCoT$_3$.

Fig. 6(a-d) The calculated thermodynamics functions for ZrCo and ZrCoX$_3$ (X= H, D and T)

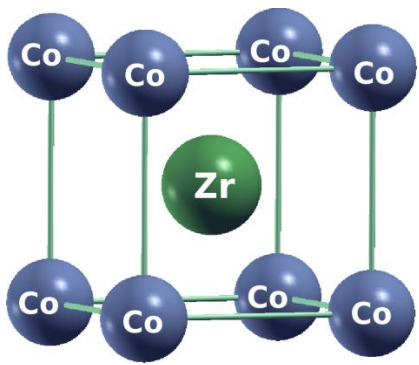    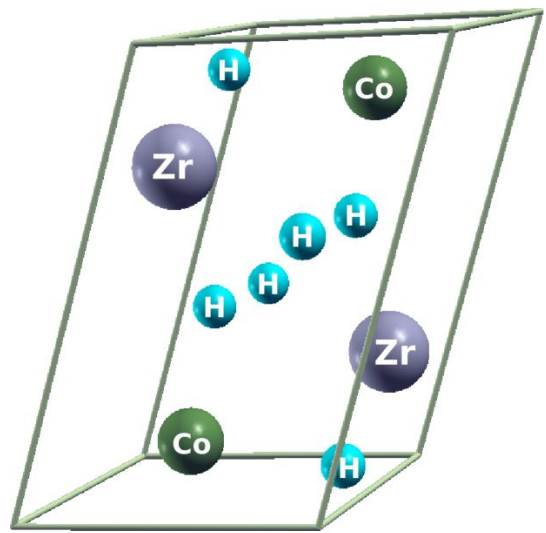

Fig. 1(a)                              Fig. 1(b)

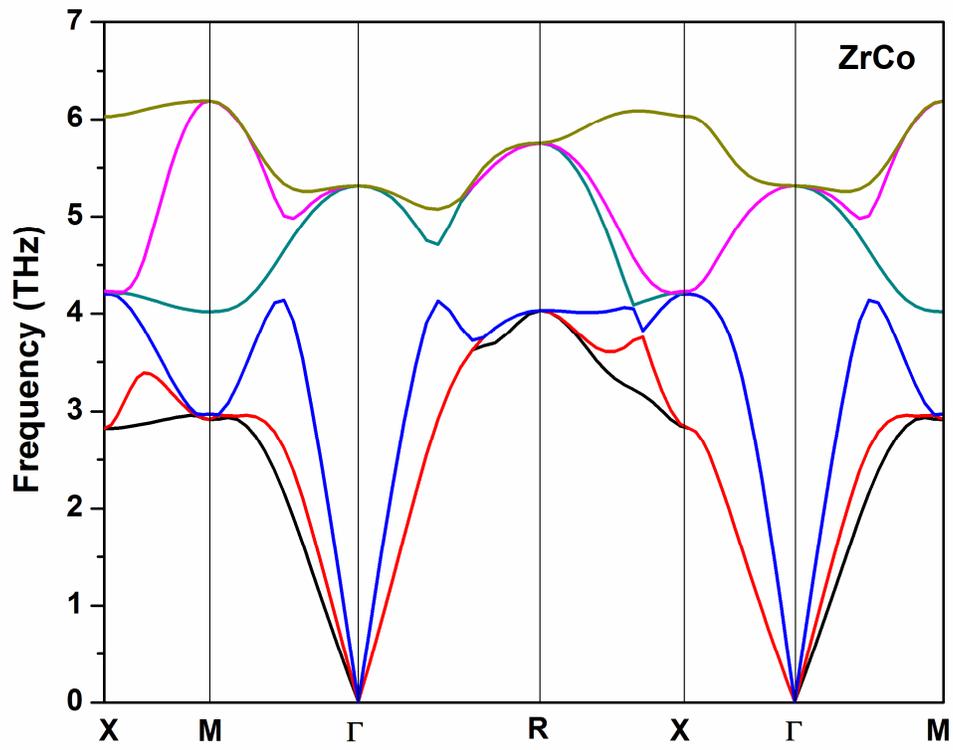

Fig. 2

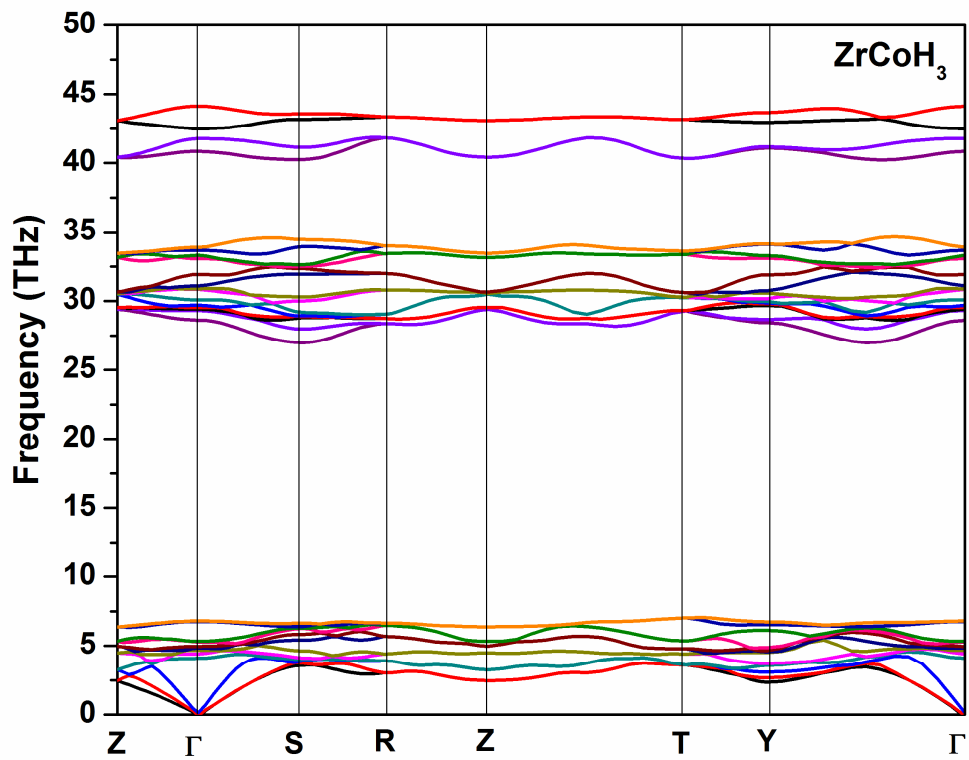

Fig. 3

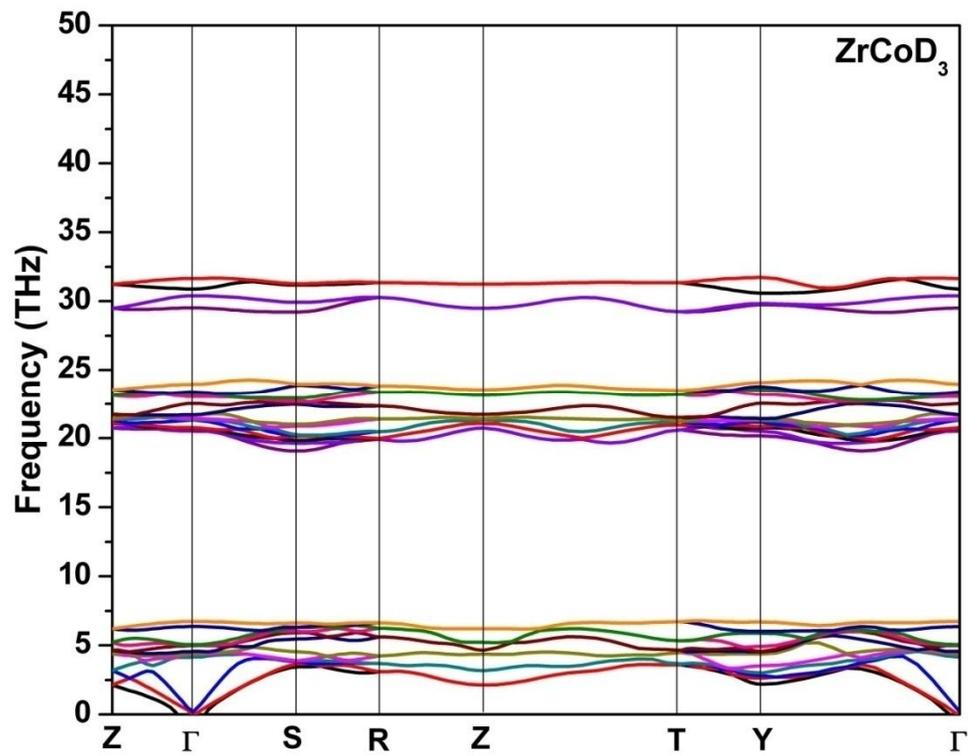

Fig. 4

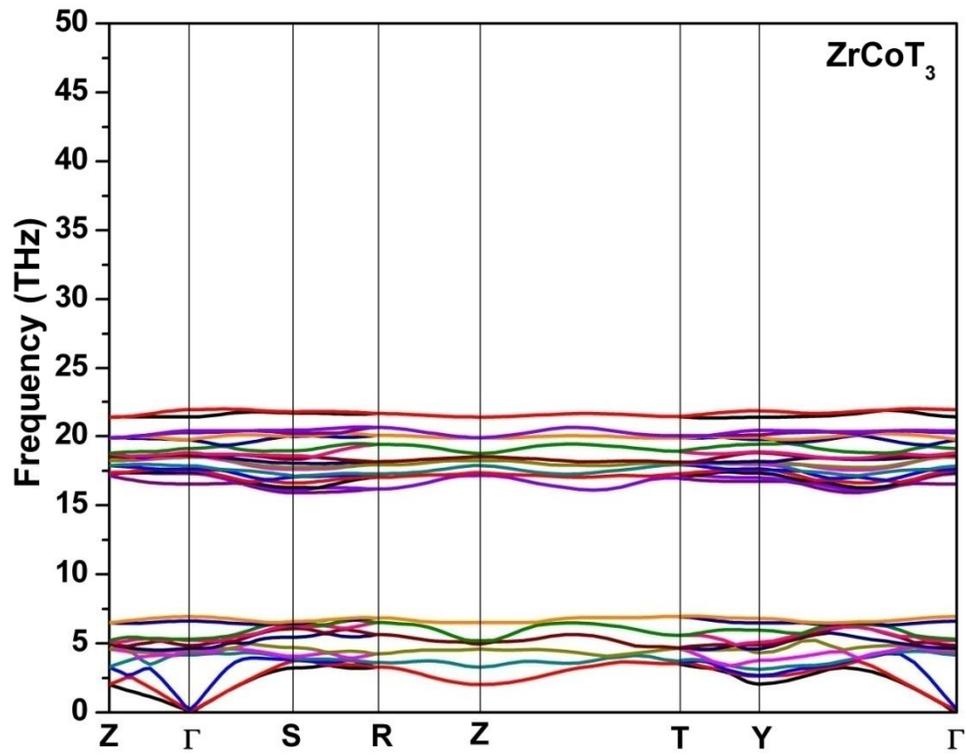

Fig. 5

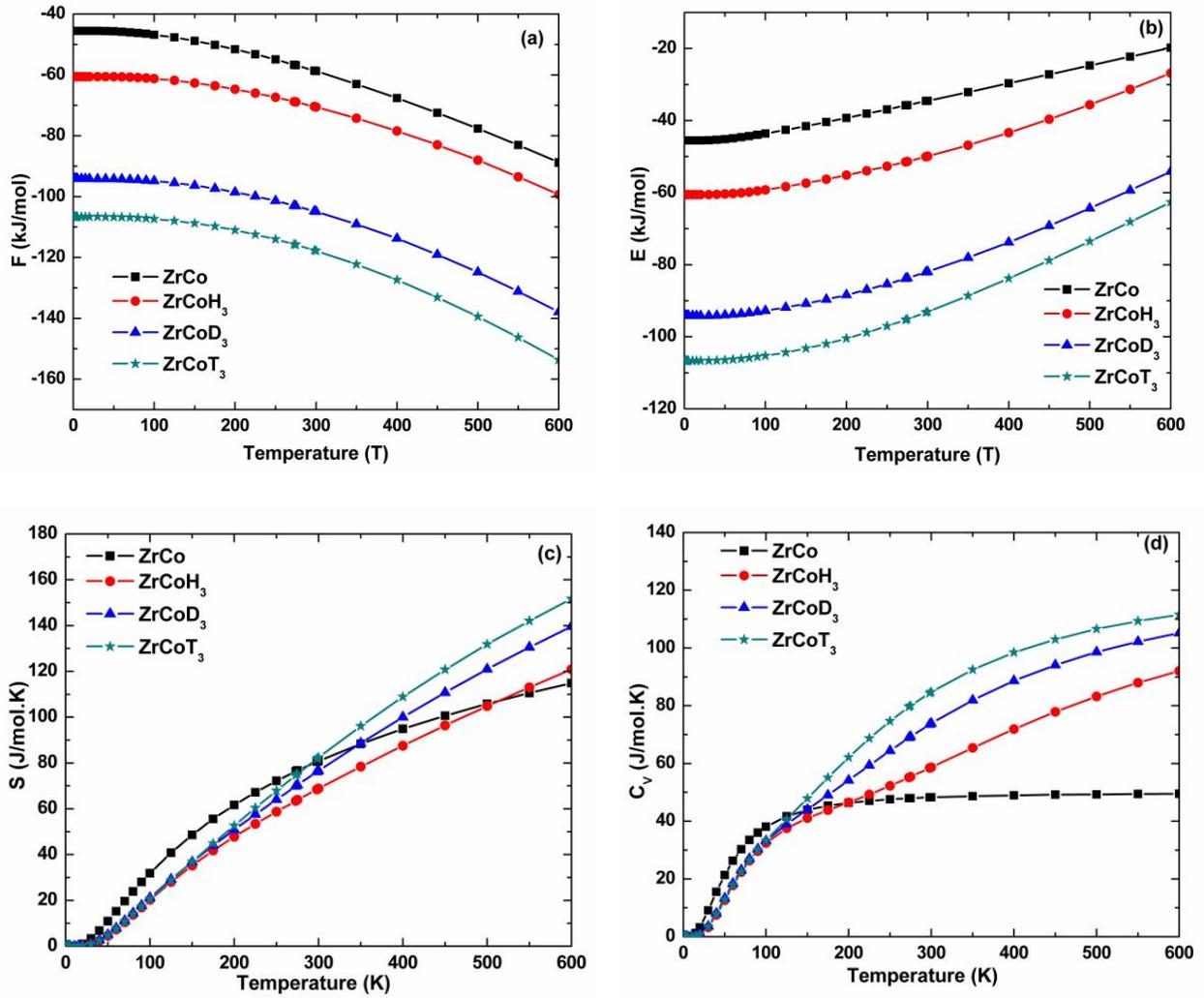

Fig. 6(a-d)